\documentclass[preprint2]{aastex}

\slugcomment{To appear in ApJ}

\shorttitle{Diffusive Shock Acceleration with Non-Resonant Streaming Instability}
\shortauthors{V.N. Zirakashvili, and V.S.Ptuskin}

\begin{document}


\title{Diffusive Shock Acceleration with Magnetic
Amplification by Non-resonant Streaming Instability in SNRs}


\author{V.N.Zirakashvili}
\affil{Pushkov Institute of Terrestrial Magnetism, Ionosphere and
Radiowave Propagation, 142190, Troitsk, Moscow Region, Russia}
\affil{Max-Planck-Institut f\"{u}r\ Kernphysik, Postfach 103980,
69029 Heidelberg, Germany}
\and
\author{V.S.Ptuskin}
\affil{Pushkov Institute of Terrestrial Magnetism, Ionosphere and
Radiowave Propagation, 142190, Troitsk, Moscow Region, Russia}



\begin{abstract}
We investigate the diffusive shock acceleration in the presence of
the non-resonant streaming instability introduced by Bell
\cite{bell04}. The numerical MHD simulations of the magnetic field
amplification combined with the analytical treatment of cosmic
ray acceleration permit us to calculate the maximum energy of
particles accelerated by high-velocity supernova shocks.
The estimates for Cas A, Kepler, SN1006, and Tycho historical
supernova remnants are given. We also found
 that the amplified magnetic
field is preferentially oriented perpendicular to the shock
front downstream of the fast shock. This explains the origin of
the radial magnetic fields observed in young supernova remnants.
\end{abstract}





\section{Introduction}

The instabilities produced by energetic particles are important
 phenomena accompanying
 the diffusive shock acceleration (Krymsky \cite{krymsky77}; Axford
et al. \cite{axford77}; Bell \cite{bell78}; Blandford and Ostriker
\cite{blandford78}) of cosmic rays in supernova remnants (SNRs). The
scattering of energetic particles both upstream and downstream of a supernova
shock is supplied by magnetic inhomogeneities existing
in the shock vicinity.
It was suggested
that this may be the result of a resonant streaming instability
that develops due to the presence of diffusive streaming of
accelerated particles (Bell \cite{bell78}). The total magnetic field
may be amplified if the energy of
unstable magnetohydrodynamic (MHD) waves becomes comparable with
the energy of the background magnetic field.

Such an amplification seems quite possible because
 quasilinear theory
of the resonant streaming instability allows it (see e.g.
McKenzie \& V\"olk \cite{mckenzie82}).

The presence of amplified magnetic fields in young SNRs is
well established now. It was indicated by the
radio observations of SNRs and was attributed to the Rayleigh-Taylor
instability of the contact discontinuity between the gas
of supernova ejecta compressed at the
reverse shock  and the circumstellar gas compressed at the
forward shock  (Gull \cite{gull75}). The radial magnetic fields
inferred from measurements of the radio-polarization in young
shell type SNRs (see e.g. Milne \cite{milne87}) may appear as this
instability develops.

However, the
discovery of thin X-ray filaments coinciding with the position of
the forward shock in young galactic SNRs (Gotthelf et
al. \cite{gotthelf01}, Hwang et al. \cite{hwang02}, Vink \&
Laming \cite{vink03}, Long et al. \cite{long03},
Bamba et al. \cite{bamba03}, Bamba et al.
\cite{bamba05}) has led to the conclusion that the magnetic field is
amplified just at the forward shock. This conclusion does not
depend on the nature of the mechanism which produces such filaments: the fast
synchrotron cooling of electrons accelerated at the forward shock
(see e.g. the analysis by Berezhko et al. \cite{berezhko02}), or
the dissipation of the MHD
turbulence and the corresponding decrease of the magnetic field, as was
suggested by Pohl et al. \cite{pohl05}.

Recently Bell \cite{bell04}, using the dispersion relation for collisionless MHD waves derived
 by Achterberg
\cite{achterberg83}, found a new regime of a non-resonant streaming
instability. In the presence of the strong electric current of accelerated particles
 a non-oscillatory purely growing MHD mode appears at spatial
 scales smaller than the
gyroradius of the particles. Bell \cite {bell04} performed MHD simulations and showed that
the magnetic field may be strongly amplified.

This instability is  investigated in more detail in our companion paper
(Zirakashvili et al. \cite{zirakashvili08}, Paper I) via numerical
MHD simulations.
Here we  combine these simulations with the analytical treatment of
diffusive acceleration
at a plane steady-state shock.
It allows us to estimate the maximum energy of the accelerated particles and to
obtain the value of the amplified magnetic field.

The paper is organized as follows.  The analytical model of  diffusive
acceleration at the plane shock is considered in Sect.2. The
MHD simulations are described in
Sect.3 and 4. The maximum energies of accelerated particles are estimated
in Sect.5.  Sect. 6 contains the discussion of obtained results.
The summary is given in the last Sect.7.

\section{Acceleration at the plane parallel shock}

We shall consider the generation of MHD turbulence and the particle acceleration
in a simple one-dimensional case and assume a steady state in the reference frame
of the shock. The applications of our results to real three-dimensional
shocks are considered in the next Sections.

The upstream plasma moves
with a velocity $u=u_1$ from $-\infty $ along the $z$ axis.
The plasma velocity downstream $u=u_2=u_1/\sigma $ drops by a factor of
$\sigma $ at  the shock front located at $z=0$.
Here $\sigma $ is the shock compression ratio.
We shall consider a parallel shock; therefore
the mean magnetic field ${\bf B}_0$ is in $z$ direction.

We shall also neglect the effects of the mean electric field
${\bf E}_0$ directed along the mean magnetic field.
The electric field  modifies the cosmic ray transport equation
(see Paper I). The isotropic part of the cosmic ray momentum distribution
$N(p,z)$ obeys the following cosmic ray transport equation
upstream and downstream of the shock:

\begin{equation}
\frac \partial {\partial z}D_{\parallel }\frac {\partial N}{\partial z}-
u\frac {\partial N}{\partial z}=0 .
\end{equation}
Here $D_{\parallel }$ is the parallel diffusion coefficient of the
energetic particles.
The cosmic ray distribution $N(p)$ is normalized as $n_{cr}=\int 4\pi p^2N(p)$,
where $n_{cr}$ is the cosmic ray number density.

The function $N$ is continuous at the shock. The boundary condition of the cosmic
ray flux conservation
at the shock front, $z=0$,  can be written as
\begin{equation}
u_1\frac p{\gamma _s}\frac {\partial N_0}{\partial p}= \left.
D_2\frac {\partial N}{\partial z}\right| _{z=+0}- \left. D_1\frac
{\partial N}{\partial z}\right| _{z=-0} \ ,
\end{equation}
where $\gamma _s=3\sigma /(\sigma -1)$, $N_0(p)$
is the distribution function at the shock, $D_1$ and $D_2$
are the parallel diffusion coefficients upstream and downstream of the
shock, respectively.

We impose an additional boundary condition $N=0$ at $z=-L$. This qualitatively
describes the escape of the highest energy particles from
a SNR with the distance $L$
being of the order of the supernova shock radius $R$.

The solution of Eq. (1) in the upstream region $z<0$ may be written as

\begin{equation}
N(z,p)=N_0(p)\frac {1-\exp \int \limits ^z_{-L}u_1dz_1/D_1(z_1,p)}
{1-\exp \int \limits ^0_{-L}u_1dz_1/D_1(z_1,p)} .
\end{equation}

\begin{figure}[t]
\includegraphics[width=7.5cm]{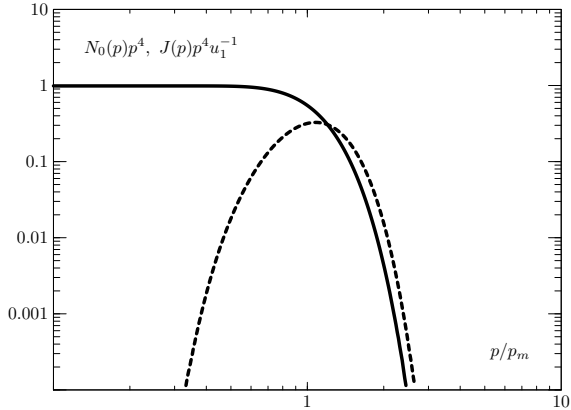}
\caption{The momentum distribution $N_0(p)$ at the shock
(solid line) and the cosmic ray flux $J(p)$ (dashed line) at
the absorbing boundary at $z=-L$. The shock compression ratio is
$\sigma =4$. }
\end{figure}

In the downstream region $z>0$ the solution is simply $N=N_0$. The
boundary condition (2) gives the ordinary differential equation for $N_0(p)$:

\begin{equation}
p\frac {\partial N_0(p)}{\partial p}=-\frac {\gamma _sN_0(p)}
{1-\exp \left( -\int \limits ^0_{-L}u_1dz/D_1(z,p)\right) } .
\end{equation}

Since the non-resonant instability produces a random magnetic field with
scales smaller than the gyroradius of the particles, the small-scale approximation
of Dolginov and Toptygin \cite{dolginov67}
can be used for the calculation of the scattering
frequency $\nu $ which determines the diffusion coefficient,
 see Paper I.
The corresponding mean free path $\Lambda =v/\nu $
is proportional to the square of the particle momentum.

If particles are scattered by the small-scale isotropic field, the
scattering frequency does not depend on the pitch-angle $\theta $
and the diffusion coefficient along the mean magnetic field is
$D_\parallel =v^2/3\nu $.  The scattering frequency $\nu =\frac
\pi 4\frac {q^2v}{p^2c^2}\int d^3k B_{\mathrm{isotr}}(k)/k$ (cf.
Paper I) is determined by the spectrum of the isotropic random
magnetic field $B_{\mathrm{isotr}}(k)$. It is normalized as
$\left< \delta B^2\right> =\int B_{\mathrm{isotr}}(k)d^3k$, where
$\left< \delta B^2\right> $ is the mean square of the random
magnetic field.

The scattering frequency depends on
the pitch-angle $\theta $ in a more general case when the
random field is statistically isotropic in the plane that is
perpendicular to the mean field direction  (see Appendix A).
Let us introduce the momentum
$p_m$ defined as:
\begin{equation}
p^2_m=\frac {3\pi }2\frac {q^2u_1}{c^3}\int \limits
^0_{-L}dzb(z) ,
\end{equation}
where the function $b(z)$ is given by the expression
\begin{equation}
b(z)=\left[ \int \limits ^{\pi /2}_0\frac {(3/8)\sin ^3\theta d\theta }
{\int d^3kB_{yy}({\bf k},z)\delta (k_z\cos \theta +k_x\sin \theta )}
\right] ^{-1} .
\end{equation}
Here $B_{yy}({\bf k})$ is the spectrum of the $y$-component of the random
magnetic field.
For the isotropic random field this function is $b=\int
d^3kB_{\mathrm{isotr}}(k)/(2k)$. The solution of Eq.(4) can then
be written as
$N_0(p)\propto p^{-\gamma _s} n_0(p/p_m)$ where the function
$n_0(s)$ with the argument $s=p/p_m$ describes the shape of the spectrum in the cut-off region:

\begin{equation}
n_0(s)=\exp \left( -\gamma _s\int \limits ^s_0\frac {ds_1/s_1}{\exp (s_1^{-2})-1}\right) .
\end{equation}

It is convenient to write down the distribution $N_0(p)$ in terms of the
cosmic ray energy flux $F_E$ at the absorbing
boundary at $z=-L$. This is the energy flux of the
highest-energy particles escaping from a SNR (the so-called run-away particles).
It may contain an essential part
$\eta _{esc}=2F_E/\rho u^3_1$ of the kinetic energy flux $\rho
u^3_1/2$, in particular when the acceleration is efficient and
the shock structure is modified by the pressure of accelerated
particles. Here $\rho $ is the plasma density.

The momentum distribution at the shock front can then be written as

\begin{equation}
N_0(p)=\frac {\eta _{esc} \rho u_1^2}{8\pi cp^{\gamma _s}
p^{4-\gamma _s}_mI}n_0(p/p_m) .
\end{equation}
Here $I=\int ^\infty _0dss^{3-\gamma _s}n_0(s)(\exp
s^{-2}-1)^{-1}$. The function $N_0(p)$ and the flux of run-away
particles at the absorbing boundary $J(p)=u_1N_0(p)/(\exp
(p^2_m/p^2)-1)$ are shown in Fig.1.

We use such a normalization of the spectrum of accelerated particles mainly
because the parameter $\eta _{esc}$ is directly related to the number density
of accelerated particles in the cut-off region. The electric current of
these particles drives the non-resonant instability (see below). It is possible to
use other parameters instead of $\eta _{esc}$, e.g. an injection efficiency of
the thermal ions at the shock front, but since its relation with the high-energy
end of the
spectrum is not so straightforward, we prefer to use $\eta _{esc}$.

In addition we have a physical reason to use this normalization.
Since generally the shock propagates in the medium with a very large
diffusion coefficient (e.g. it is higher than $10^{28}$
cm$^2$s$^{-1}$ in the interstellar medium, cf. Berezinskii et al.
\cite{berezinsky90}), the accelerated particles are effectively
scattered only near the shock front where the level of
the self-excited MHD turbulence is high. The accelerated particles
with the maximum energies run away from this region to the outer
space. Thus the acceleration at the plain shock with the absorbing
boundary considered in this paper simulates the real situation
when the acceleration by the three-dimensional supernova shock is
considered.

The possible effect of the shock modification by the cosmic
ray pressure was not taken into account in Eq.(4).
The modification may strongly change the spectrum of
particles and make it concave at low energies (see e.g. Malkov \& Drury \cite{malkov01} for
a review). However, even in this case the
cut-off of the spectrum is described accurately by Eq. (7).
The total shock transition that
includes the precursor created by the pressure of accelerated particles and
the thermal sub-shock may be considered as a sharp
discontinuity for the cosmic ray particles with the highest energies.

The calculation of the diffusive electric current of accelerated
particles with the use of Eq. (8) gives
\begin{equation}
j_d(z)=\frac {\eta _{esc} \rho u_1^3q}{2cp_mI} \int \limits ^\infty
_0dss^{2-\gamma _s} \frac {\exp \left( \frac
{a(z)}{s^2a(0)}\right) }{\exp (s^{-2})-1}n_0(s) ,
\end{equation}
where $a(z)=\int \limits ^z_{-L}dz_1b(z_1)$.

We shall use the values $\gamma _s=4$ and $I=1/4$ below. This
corresponds to an unmodified strong shock. However, the accepted
value of $\gamma _s$
does not strongly differ from $\gamma _s\sim 3.7-3.8$, typical for
the moderately modified shocks.

\begin{figure}[t]
\includegraphics[width=7.5cm]{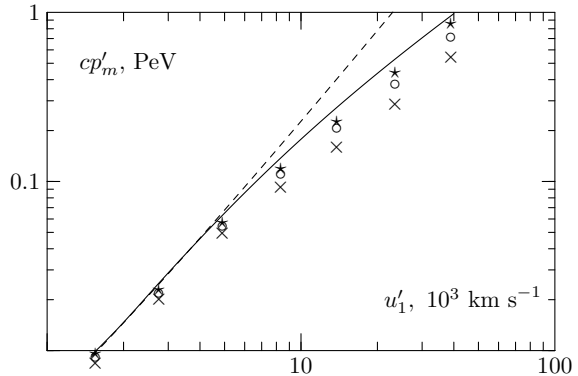}
\caption{The dependence of the normalized maximum momentum $p'_m$
on the normalized shock velocity $u'_1$. The results were obtained
using the resolution $64^3$ (crosses), $128^3$ (open circles) and
$256^3$ (stars). The analytical approximation and the upper limit
according to Eq. (19) are shown by the solid and dashed lines,
respectively. }
\end{figure}

\section{Modeling of the non-resonant instability with diffusive shock acceleration}

We can now model the magnetic field amplification in the vicinity of
the shock which accelerates particles. We shall seek the steady state
solution for the spectrum of accelerated particles and the time-averaged
MHD spectra.

Since the shock velocity is much higher than the phase velocities
of MHD waves and the turbulent velocities of the plasma upstream of
the shock, we can model the dependence of the MHD spectra on the
distance from the shock front via the investigation of the temporal
evolution of the instability in the simulation box moving with the
gas flow in the
direction of the shock front. This strongly reduces the size of
the simulation box in the $z$-direction and permits to obtain the
numerical results with  a good numerical resolution.
The computation time is also
significantly reduced.

We shall assume that the shock propagates in a medium with  the density
$\rho _0$, the gas pressure $P_0$ and the Alfv\'en velocity $V_a=B_0/\sqrt{4\pi \rho _0}$.

The details of the numerical method were given in Paper I.

The dimensionless time $\tilde{t}$, the space coordinate $\tilde{z}$ and
the velocity $\tilde{u}$ are determined as $\tilde{t}=tV_ak_0$,
$\tilde{z}=k_0z$, $\tilde{u}=u/V_a$. Here $k_0$ is the wavenumber
that corresponds to the real size $2\pi /k_0$ of the box. The
dimensionless density $\tilde{\rho }$ and the electric current $J$ can
be expressed via the magnetic field $B_0$ and the Alfv\'en velocity $V_a$ as
$\tilde{\rho }=4\pi \rho V_a^2/B_0^2$ and $J=4\pi j/ck_0B_0$.

The real size of the simulation box is small in comparison with
the characteristic scale of the spatial distribution of the
accelerated particles upstream of the shock. At $\tilde{t}=0$ the
box is placed at $z=-L$, where the initial background random
magnetic field corresponding to the isotropically distributed
Alfv\'en waves with the one dimensional spectrum $\propto k^{-1}$
and the amplitude $\left< \delta B^2\right> ^{1/2}=0.09B_0$  is
preset. We use the gas adiabatic index  $\gamma =5/3$ and the
parameter $\beta =1$. Here $\beta = 4\pi P_0/B_0^2$.

The box moves with the mean flow speed $u_1$ towards the shock.
The MHD equations which include the  Lorentz force produced by the
electric current of accelerated particles are solved numerically
in the three dimensions (see Paper I for detail).

Let us fix the momentum $p_m$. At every
instant of time the diffusive electric current of accelerated particles,
 that drives
the instability, is calculated according to Eq. (9). It may be
rewritten in dimensionless units as

\begin{equation}
J(\tilde{t})=J_0 \int \limits ^\infty _0\frac {ds}{s^2} \frac
{\exp \left( gs^{-2}\int
^{\tilde{t}}_0d\tilde{t}'b(\tilde{t}')\right) } {\exp
(s^{-2})-1}n_0(s) .
\end{equation}
Here $g$ is some arbitrary dimensionless constant. We change the
integration over $dz_1$ in Eq. (9) to the integration over time
$dt'=dz_1/u_1$ in Eq. (10). The dimensionless current $J_0$ can then be
written in terms of the physical parameters as

\begin{equation}
J_0=\frac {2\eta _{esc} u_1^2}{V_a^2}\left( \frac {2gV_a}{3\pi
c}\right) ^{1/2}.
\end{equation}

The simulation is performed up to the point in the dimensionless time
$\tilde{t}_m$ at which the value of the integral $\int
^{\tilde{t}}_0d\tilde{t}'b(\tilde{t}')$ reaches $g^{-1}$. This
corresponds to the box arrival to the position of the shock.
It means that we have found
the size $L$ for
the given maximum momentum $p_m$.  Or,
inversely, the  momentum $p_m$ can be found from Eq. (5):

\begin{equation}
\frac {p_m}{mc}=\tilde{t}_m^{-1}\left( \frac {3\pi
V_a}{2gc}\right) ^{1/2} \frac {qB_0L}{mc^2}.
\end{equation}

Several runs were performed to scan the broad range of physical
parameters. We used the value $J_0=64$ and several values of $g>1$ and
$J_0=64g^{-1}$ for $g<1$. This choice limits the characteristic
wavenumber of the generated magnetic field to the value about 4
and therefore the characteristic scale of magnetic field is
smaller than the size of the box.

\begin{figure}[t]
\includegraphics[width=7.5cm]{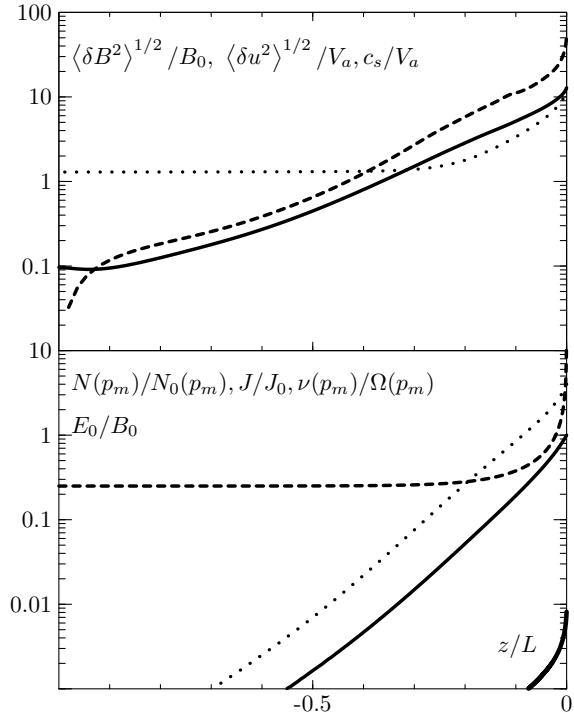}
\caption{Dependence of the physical quantities on $z$. {\it Top}:
random magnetic field (solid line), turbulent velocity (dashed
line) and sonic velocity (dotted line). {\it Bottom}: momentum
distribution $N(p_m)$ (solid line), diffusive electric current $J$
(dashed line), scattering frequency $\nu (p_m)$ normalized to the
gyrofrequency $\Omega (p_m)=qB_0/p_m$ (dotted line) and the mean
electric field $E_0$ (thick solid line). The calculations were
performed for the normalized shock speed $u'_1=4900$ km s$^{-1}$.
}
\end{figure}
\begin{figure}[t]
\includegraphics[width=7.5cm]{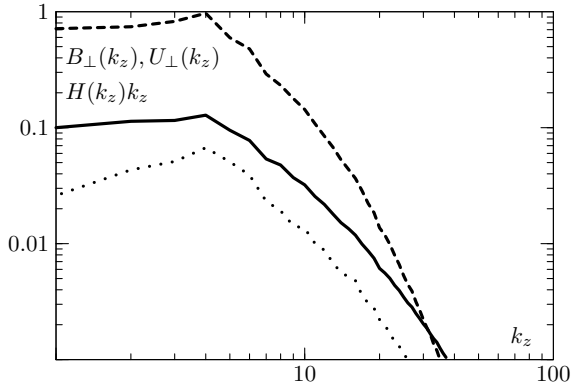}
\caption{The one-dimensional spectra of the perpendicular component of
 the random magnetic field $B_\perp (k_z)$ (solid
line), turbulent velocity $U_\perp (k_z)$ (dashed line) and magnetic
helicity $H(k_z)$ (dotted line) at the shock front. The calculations
were performed for the normalized shock speed $u'_1=4900$ km
s$^{-1}$. All spectra are normalized to the mean square of the
perpendicular component of the random magnetic field. }
\end{figure}
\begin{figure}[t]
\includegraphics[width=7.5cm]{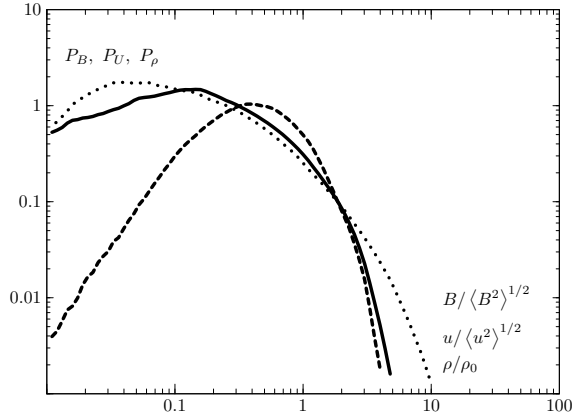}
\caption{PDF of the magnetic field $P_B(B)$ (thin solid line),
turbulent velocity $P_U(u)$ (thin dashed line) and density
$P_{\rho }(\rho )$ (thin dotted line) at the shock front.
The numerical calculations were
performed for the normalized shock speed $u'_1=4900$ km s$^{-1}$.
}
\end{figure}

It is convenient to present the numerical results in terms of
the normalized shock velocity $u'_1$ defined as

\begin{equation}
u'_1=u_1\ \left( \frac {V_a}{\mbox{10 km s}^{-1}}\right) ^{-3/4}
\left( \frac {\eta _{esc}}{0.05}\right) ^{1/2}
\end{equation}
and the normalized maximum momentum $p'_m$, given by

\begin{equation}
p'_m=p_m\ \left( \frac {V_a}{\mbox{10 km s}^{-1}}\right) ^{-1/2}
\left( \frac {L}{\mbox{1 pc}}\right) ^{-1} \left( \frac {B_0}{5\mu
\mbox{G}}\right) ^{-1}.
\end{equation}

We normalize the shock velocity and maximum momentum using the
parameter value
 $\eta _{esc}=0.05$. This value corresponds in particular to
the case when the energetic spectrum of CR particles at the shock front
is a power-law with the exponent $\gamma _s-2= 2$ and
the total CR pressure equals  $0.5\rho _0u^2_1$.
The dependence of the normalized maximum momentum $p'_m$ on the
shock velocity $u'_1$ is shown in Fig.2.


\begin{table*}[t]
\begin{center}
\caption{Results of the modeling of the non-resonant instability
with diffusive shock acceleration }
\begin{tabular}{llllllll}
\tableline
$u_1',\ 10^3\ \mathrm{km/s}\tablenotemark{a}$& 1.55 & 2.76& 4.90& 8.25& 13.9& 23.3& 39.2\\
$p_m',\ \mathrm{TeV}/c     \tablenotemark{b}$& 9.46 & 22.7& 56.4& 120 & 227 & 443 & 854 \\
$\left< \delta B^2\right> ^{1/2}/B_0    \tablenotemark{c}$& 1.36 & 4.68& 12.6& 24.2& 44.8& 87.3& 187 \\
$r_b                       \tablenotemark{d}$& 0.77 & 0.71& 0.77& 0.85& 0.85& 0.84& 0.82\\
$\left< \delta u ^2\right> ^{1/2}/V_a   \tablenotemark{e}$& 5.15 & 23.6& 49.9& 94.1& 166 & 361 & 808 \\
$c_s/V_a                   \tablenotemark{f}$& 1.42 & 3.70& 11.9& 31.6& 68.4& 139 & 289 \\
$(V_a/10\ \mathrm{km}\ \mathrm{s}^{-1})^{1/4}(\eta
_{esc}/0.05)^{-1/2}\nu (p_m)/\Omega (p_m)    \tablenotemark{g}$
                                             & 1.44 & 2.63& 3.95& 4.96& 7.21& 12.6& 22.4\\
$10^2(V_a/10\ \mathrm{km}\ \mathrm{s}^{-1})^{-1}E_0/B_0
\tablenotemark{h}$
                                             & 0.016& 0.14& 0.74& 2.86& 10.6& 43.7& 209 \\
$10^2(V_a/10\ \mathrm{km}\ \mathrm{s}^{-1})^{-1/2}q\phi /p_mc
\tablenotemark{i}$
                                             & 0.348& 1.03& 2.03& 3.30& 5.00& 7.52& 11.0\\
$10^2D(p_m)/u_1L   \tablenotemark{k}$& 9.2&    6.8&  6.3&  6.3&  4.9&  3.3&  2.1\\
$(V_a/10\ \mathrm{km}\ \mathrm{s}^{-1})^{-1/4}(\eta
_{esc}/0.05)^{1/2}k_0p_mc/qB_0             \tablenotemark{l}$
                                             & 0.194& 1.09& 6.15& 14.6& 34.8& 82.7& 197 \\
\tableline
\end{tabular}


\tablenotetext{a}{Normalized shock velocity $u_1'$.}
\tablenotetext{b}{Normalized momentum $p_m'$ of accelerated
protons.} \tablenotetext{c}{Ratio of rms of the random magnetic
field to the mean magnetic field.} \tablenotetext{d}{ Ratio of rms
of the random parallel component $\left< (\delta B_z^2)\right>
^{1/2}$ of the magnetic field  to the corresponding perpendicular
component $\left< (\delta B^2_x+\delta B^2_y)/2\right> ^{1/2}$}
\tablenotetext{e}{Ratio of rms of the turbulent velocity to the
Alfv\'en velocity. } \tablenotetext{f}{Ratio of the sound velocity
to the Alfv\'en velocity.} \tablenotetext{g}{Ratio of the
scattering frequency $\nu (p_m)$ to the gyrofrequency $\Omega
(p_m)=qB_0/p_m$.} \tablenotetext{h}{Ratio of the mean electric
field to the mean magnetic field. } \tablenotetext{i}{ Ratio of
the potential energy of the particle $q\phi $ corresponding to the
electric potential $\phi $ to the energy $p_mc$. }
\tablenotetext{k}{Ratio of the diffusion length $D(p_m)/u_1$ and
$L$} \tablenotetext{l}{Product of the wavenumber $k_0$ and the
gyroradius $p_mc/qB_0$. The wavenumber determines the real size
$2\pi /k_0$ of the simulation box.} \tablecomments{All quantities
from the lines c-k are calculated at the shock front.}

\end{center}
\end{table*}


The spatial dependence of the random magnetic field, the turbulent
velocity, the sonic velocity, the momentum distribution $N(p_m)$,
the diffusive electric current, the scattering frequency $\nu (p_m)$
 and the
mean electric field $E_0$
obtained for the normalized speed $u'_1=4900$ km s$^{-1}$  are shown in Fig.3.

The spectra of the magnetic field, the plasma velocity and the magnetic
helicity at the shock front are shown in Fig.4.

For physical applications it is useful
to know the probability distributions functions (PDFs) of different
physical quantities. The PDF of the random magnetic field, the turbulent velocity and the gas
density obtained in our simulations at the shock front are shown in Fig.5.

As it is seen in Fig. 3 the diffusive electric current sharply increases near the shock.
As a result the perpendicular components of the turbulent velocity also increase
and the kinetic energy of the random motions just upstream of the
shock is an order of magnitude larger than the magnetic energy.

It is important that the accelerated
particles are concentrated in the shock vicinity. This justifies
the use of the planar geometry even for real three-dimensional shocks.
The amplification of the MHD turbulence takes
place in the broader region where the diffusive electric current
of run-away particles is not small.

On the whole our modeling corresponds to the following physical picture.
Let us consider a volume element at some distance from the
supernova. Shortly after the explosion the run-away particles
reach the volume and drive the streaming instability. Our numerical
modeling shows how the magnetic fluctuations are amplified in the
volume element. For simplicity we assumed that the electric
current was constant at all times after the explosion before the shock arrival.
This assumption is strictly valid for a steady state plane shock (see Fig.3) and only
qualitatively valid for three-dimensional shocks.
The dimensionless
time $\tilde{t}_m$ in our calculations corresponds to the supernova remnant age $T$.
If the shock radius $R$ increases as
$R\propto t^{m_0}$, where $m_0$ is the expansion parameter, then
$T=m_0R/u_1$ and we should use the relation $L=m_0R=u_1T$ for the
parameter $L$ in Eqs (12) and (14).  Note, that
the electric current increases according to our plane shock modeling
when the shock comes close to
the volume element.

A summary of the numerical results obtained for different normalized
shock velocities $u_1'$ is given in Table 1.

\begin{figure}[t]
\includegraphics[width=7.5cm]{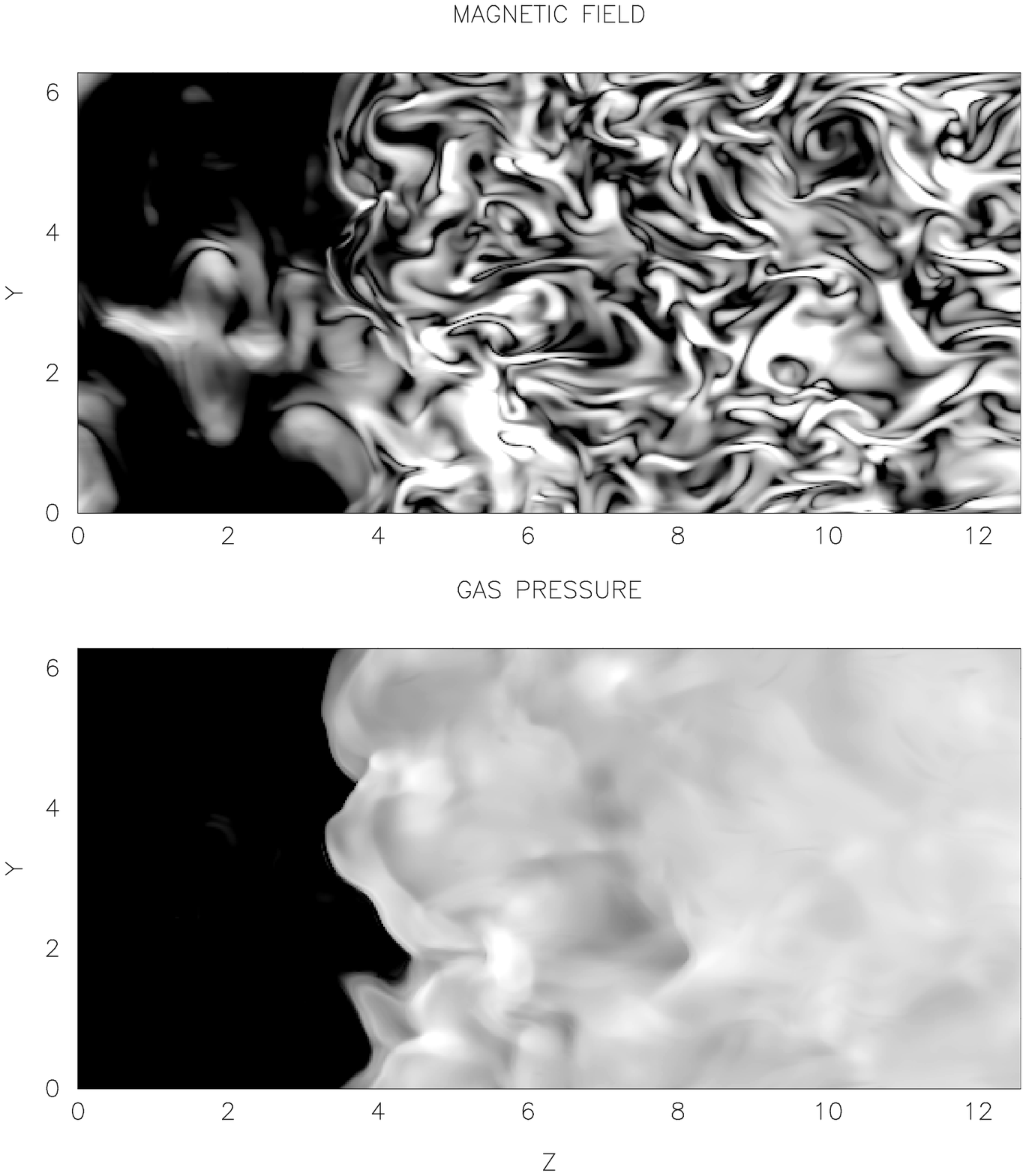}
\caption{ Slices of the magnetic field strength (top) and
gas pressure (bottom) through the center of the box obtained at
$\tilde{t}=0.12$ in the shock transition region and downstream of
the shock. The scaling of the magnetic field strength is
logarithmic between $12.6B_0$ (black) and $126B_0$ (white). The
scaling of the gas pressure is also logarithmic between
$10^3B^2_0/4\pi $ (black) and $10^5B^2_0/4\pi $ (white). The
calculations were performed for a
 shock propagating with a speed $u_1=3000$ km s$^{-1}$ in the medium with $V_a=10$ km s$^{-1}$. }
\end{figure}

\section{MHD modeling in the shock transition region and downstream of the shock}

The density fluctuations arising during the development of
the non-resonant instability, when the expanding magnetic
spirals collide each other (cf. Paper I), play a crucial role in the shock
transition region. It is well known, that the propagation of the
shock in the medium even with small density fluctuations is
accompanied   by  shock front distortions and the appearance    of
 vortex plasma motions downstream of the shock (see e.g.
Kontorovich \cite{kontorovich59}, McKenzie \& Westphal
\cite{mckenzie68}, Bykov \cite{bykov82}). These motions can
amplify the magnetic field downstream of the shock even in the
absence of accelerated particles. This effect was recently
observed by Giacalone \& Jokipii \cite{giacalone07} in their 2D MHD
numerical simulations. The results of low resolution three dimensional
calculations for the MHD evolution of an adiabatic supernova
remnant in a non-uniform and turbulent interstellar medium
 were presented by Balsara et al. \cite{balsara01}.

\begin{figure}
\includegraphics[width=7.5cm]{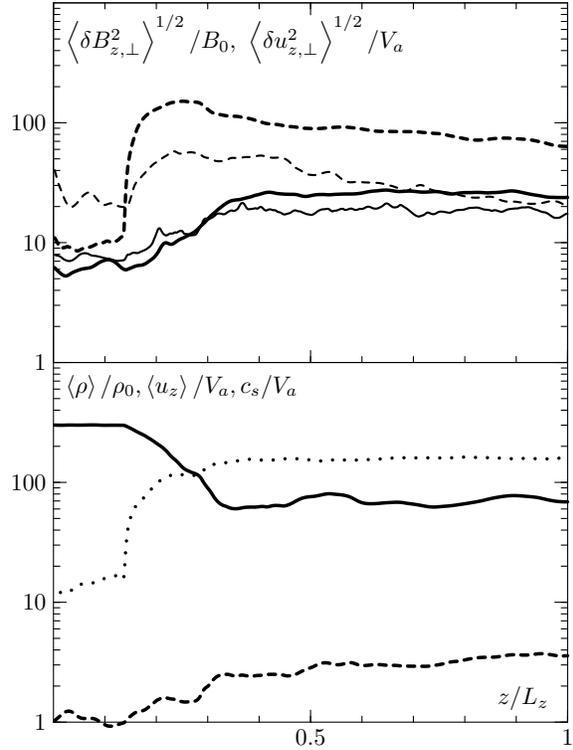}
\caption{Dependence of the physical quantities on $z$ in the shock
transition region. {\it Top}: random $z$- and perpendicular
magnetic field components (normal solid and thin solid lines
respectively), $z$- and perpendicular components of the  turbulent
velocity (thick and thin dashed lines respectively). {\it Bottom}:
sonic velocity (dotted line),
 mean gas density (dashed line), mean gas velocity (solid line)
in $z$ direction . The calculations were performed for the
 shock propagating with a speed $u_1=3000$ km s$^{-1}$ in the medium with $V_a=10$ km s$^{-1}$. }
\end{figure}

We investigated this phenomenon and studied the evolution of MHD turbulence
in the downstream region. The simulation box with the size
$2\pi \times 2\pi \times 4\pi $, stretched in the $z$ direction was used.

The gas flows with the shock velocity $u_1$ along the $z$-axis
into the box from its left boundary. At the initial moment of time
the flat shock front is located at $z=2\pi $. The plasma
compression and heating downstream are taken in accordance with
the Rankine-Hugoniot  conditions. During the simulation the plasma
magnetic field, density and pressure are prescribed at the left
boundary in accordance with the spatially periodic numerical
solution, found in the previous Section. The plasma leaves the
system at the right boundary, where the homogeneous boundary
conditions are prescribed. The periodic boundary conditions in the
perpendicular directions are used. We performed 3D MHD simulation
with $256^2\times 512$ cells using the numerical method described
in Sect.3. The electric current of accelerated particles was
switched off.

\begin{figure}[t]
\includegraphics[width=7.5cm]{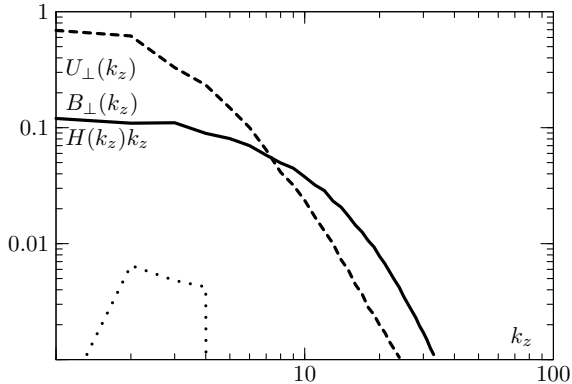}
\caption{The one-dimensional spectra of the perpendicular component of
 the random magnetic field $B_\perp (k_z)$ (solid
line), turbulent velocity $U_\perp (k_z)$ (dashed line) and magnetic
helicity $H(k_z)$ (dotted line) downstream of the shock.
 The calculations were performed for a
 shock propagating with a speed $u_1=3000$ km s$^{-1}$ in the medium
 with $V_a=10$ km s$^{-1}$. All spectra are normalized to the mean square of the
perpendicular component of the random magnetic field. }
\end{figure}

The obtained numerical results at the shock velocity $u_1=3000$ km s$^{-1}$
and the Alfv\'en velocity $V_A=10$ km s$^{-1}$ are shown in Fig.6
and Fig.7. We used the numerical solution with the normalized shock
velocity $u'_1=4900$ km s$^{-1}$ described in the previous
Section. The corresponding parameter $\eta _{esc}=0.14$.

The real size of the
box in $z$ direction $L_z=4\pi /k_0$ is related to the distance to
the absorbing boundary $L$
as $L_z=4\pi LV_a/(u_1\tilde{t}_m)$. The value $\tilde{t}_m=1.02$
corresponds to the numerical solution with the normalized shock velocity
$u'_1=4900$ km s$^{-1}$ described in the previous Section.

The slices
 of the magnetic field strength and the gas pressure in the $YZ$ plane are shown in the
top and bottom panels of Fig.6 respectively. The strong
distortions of the shock front and the sonic waves propagating in the
downstream region are clearly seen in the bottom panel. The shock
front is shifted to the right due to the interaction with the fluid
elements which have the enhanced density. The magnetic structures
downstream are stretched along the direction of the flow (top
panel).

The dependence of the
plasma parameters averaged in the perpendicular $XY$ plane on $z$-coordinate
 is shown in Fig.7. Strong fluctuations of the plasma motions with amplitude of the order
of one third of the shock velocity $u_1$ exist downstream. It is
remarkable that these random
motions occur mainly in $z$ direction.
The magnetic field is also stretched in this direction.
The $z$-component of the random magnetic field is a factor of 1.4 larger than
the perpendicular components
downstream of the shock.

\begin{figure}[t]
\includegraphics[width=7.5cm]{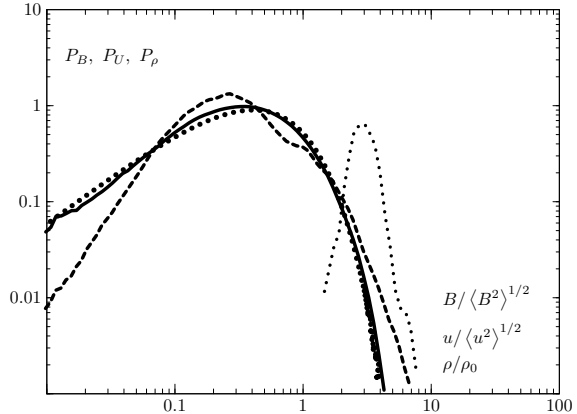}
\caption{PDF of the magnetic field $P_B(B)$ (thin solid line),
turbulent velocity $P_U(u)$ (thin dashed line) and density
$P_{\rho }(\rho )$ (thin dotted line) downstream of  the shock. The
analytical approximation for PDF of the magnetic field (23) (thick
dotted line) is also shown. The calculations were performed for a
 shock propagating with a speed $u_1=3000$ km s$^{-1}$ in the medium
 with $V_a=10$ km s$^{-1}$.
}
\end{figure}

The spectra of the magnetic field, the plasma velocity and the magnetic
helicity downstream of the shock are shown in Fig.8. Note, that the
magnetic helicity is not zero in this region.

PDF distributions of the random magnetic field, turbulent velocity and plasma density
obtained downstream of the shock are shown in Fig.9.

The numerically calculated magnetic compression ratio $\sigma _B$ is
about 3.0 in this simulation with the shock compression ratio $\sigma
$ close to 4. This value of $\sigma _B$ is close to one that is
expected if the isotropic random magnetic field is compressed at
the shock front: $\sigma _B=\sqrt{(2\sigma ^2+1)/3}$. Since in the
real situation the pressure of accelerated cosmic ray particles
should be taken into account, the shock compression will be
higher. We found that the use of the adiabatic index $\gamma =4/3$
with the corresponding shock compression ratio about 6 results in a
magnetic compression factor $\sigma _B$ close to 3.5, that is
smaller than the expected value $\sigma _B\sim 5$. Probably this is due to
 the numerical dissipation of the field. It is
necessary to have  a better numerical resolution in the last case
since the stronger compression of the field at the shock front
results in the stronger numerical dissipation. We obtained the
close value of $\sigma _B=2.6$  in the   simulation
with $\gamma =5/3$ and the lower numerical resolution $256\times 128^2$. This
demonstrates that the resolution $512\times 256^2$ is good enough
for the simulation of the shock with the compression ratio 4. We found
that the corresponding values of $\sigma _B$
 differ significantly from each other in the case of the higher shock compression ratio.

\section{Analytical estimate of the maximum energy of accelerated particles
in SNR}

Now we can estimate analytically the maximum particle energy
achieved in the process of acceleration. We shall use Eq. (5) and
assume that the random field is isotropic and concentrated at the
wavenumber $k$, so that $b=B_r^2/(2k)$ where the rms of the random
magnetic field $B_r=\left< \delta B^2\right> ^{1/2}$. Let us
change the integration on $dz$ to the integration on time
$dt=dz/u_1$ in Eq. (5). The evolution of the amplified magnetic
field can be approximately described as
\begin{equation}
B_r(t)=\left\{ \begin{array}{ll}
B_{b}\exp (\gamma _{\max }t), \ t<t_1=\gamma _{\max }^{-1}\ln \frac {2B_0}{B_b} \\
2B_0(1+\gamma _{\max }(t-t_1)), \ t>t_1 \end{array} \right.
\end{equation}
Here $B_b$ is the initial value of the amplified field. At the first
stage $t<t_1$ the field is amplified exponentially in time with
the maximum growth rate $\gamma _{\max}=j_dB_0/2c\rho _0V_a$ (cf. Paper I).
The
corresponding wavenumber $k$ is $k=2\pi j_d/cB_0$ at this stage.
At the second stage
$t>t_1$ the field is amplified linearly in time.
The wavenumber $k$ may be found from the expression $kB_r=4\pi
j_d/c$. This dependence of random magnetic field on time
is in a qualitative agreement with the simulations of
the non-resonant instability (cf. Paper I).

The integration of Eq. (5) gives

\begin{equation}
p_m^2=\frac {3\pi }{4}\frac {q^2u_1^2B_0^2V_a}{\gamma _{\max
}^2c^3}
\left\{ \begin{array}{ll}
B_r^2/(2B_0^2), \ t_m<t_1\\
1+(1+\gamma _{\max }(t_m-t_1))^4 , \\
t_m>t_1.
\end{array} \right.
\end{equation}
Here $t_m=L/u_1$. The electric current $j_d$ that drives the
instability is constant over the whole region $-L<z<0$
except the narrow zone near the shock (see Fig.3). The input of
this zone was neglected  in Eq. (16). This assumption
introduces a relatively small error into the calculation of $p_m$,
since Eq. (16) is an implicit equation for $p_m$. The growth rate
$\gamma _{\max }$ is inversely proportional to $p_m$ (see Eq. (17)
below). Thus $p_m$ remains in Eq. (16) only through $\gamma _{\max }$
after the parenthesis. This produces an error in the evaluation of
$t_m$ and $p_m$ that is not larger than 10 percent.

Using Eq.
(9) we estimate $j_d=\eta _{esc}q\rho u_1^3/(2cp_m)$ and
\begin{equation}
\gamma _{\max }=\frac {\eta _{esc}}4\frac {qu_1^3B_0}{V_ac^2p_m}.
\end{equation}

From these two equations we finally obtain the value of the amplified field

\begin{equation}
B_r=2B_0
\left\{ \begin{array}{ll}
u_1^2/u_*^2, \ u_1<u_* \\
\left( 2\frac {u_1^4}{u_*^4}-1\right) ^{1/4}, \ u_1>u_*
\end{array} \right.
\end{equation}
and the maximum energy of accelerated particles
\begin{equation}
p_mc=\frac {\eta _{esc}qu_1^2B_0L}{4cV_a}
\left\{ \begin{array}{ll}
\ln ^{-1}\left( \frac {2B_0u_1^2}{B_bu_*^2}\right) , u_1<u_*\\
\left[ \ln \left( \frac {2B_0}{B_b}\right) -1 +\left( 2\frac
{u_1^4}{u_*^4}-1\right) ^{1/4} \right] ^{-1}, \\
u_1>u_*
\end{array} \right.
\end{equation}
where the velocity $u_*=(24\pi cV_a^3/\eta _{esc}^2)^{1/4}$.

Using the first line of expression (19) for an arbitrary shock velocity
$u_1$ we obtain the upper limit of the maximum energy.
The growth rate (17) calculated with this maximum energy is high enough to
provide the considerable magnetic field growth during the age of
a supernova remnant
(to be compared with Bell \cite{bell04}). The upper
limit and the maximum energy (19) for $\ln \left(
2B_0/B_b\right)=5$ are shown in Fig.2. We found that for
$u_1>>u_*$  Eq. (18) gives the amplified field a factor of 2
smaller than the numerical results (see Table 1). It is because we
neglected the growth of the diffusive current $j_d$ in the narrow zone
just adjacent to the
shock in our analytical estimates.
However, this does not influence the estimate of the
maximum energy (19).


\begin{table*}[t]
\begin{center}
\caption{Maximum energies and amplified magnetic fields in historical SNRs.  }
\begin{tabular}{lcccccccccccccc}
\tableline\tableline
 & $T\tablenotemark{a}$ &
$u_1\tablenotemark{b}$ & $n_H\tablenotemark{c}$
&$B_0\tablenotemark{d}$&$u_1'\tablenotemark{e}$
&$u_1'\tablenotemark{e}$&$u_1'\tablenotemark{e}$&$p_mc\tablenotemark{f}$&$p_mc\tablenotemark{f}$&
$p_mc\tablenotemark{f}$&
$B_d\tablenotemark{g}$&$B_d\tablenotemark{g}$
&$B_d\tablenotemark{g}$& $B_d^{obs}\tablenotemark{h}$ \\ \tableline
$\eta _{esc}$& & & & &0.01&$0.05$&$0.14$&0.01&$0.05$& $0.14$&0.01&$0.05$&$0.14$& \\
Tycho& 435 & 4500& 0.3 & 5.0 &1360&3040&5090&19&76& 170&21&110& 260 & 300\\
SN1006     &1000 &4300& 0.1 & 5.0 &860&1920&3220&22& 100&240 &8.4&43&120 & 140 \\
Kepler    & 400 &5300& 0.35& 5.0 &1670&3800&6360&30&115&250&33&160 &350& 215 \\
Cas A& 330 &5200&3.0& 10.0&2220&4960&8300&63&230& 500 &120&510& 980& 485\\
\tableline
\end{tabular}

\tablenotetext{a}{Age of SNR, yr} \tablenotetext{b}{
Forward shock velocity, km s$^{-1}$} \tablenotetext{c}{
Hydrogen number density of the circumstellar medium, cm$^{-3}$}
\tablenotetext{d}{Magnetic field strength in the circumstellar
medium, $\mu $G } \tablenotetext{e}{Normalized shock velocity,
km s$^{-1}$ } \tablenotetext{f}{Calculated maximum energy
$p_mc$, TeV} \tablenotetext{g}{Calculated downstream magnetic
random field strength, $\mu $G } \tablenotetext{h}{Downstream magnetic field
determined from the thickness of X-ray filaments, $\mu $G}
\tablecomments{The numbers from the columns $b,c$ are taken from Vink
\cite{vink06}, the numbers in the last column $h$ are according to
V\"olk et al. \cite{voelk05}}

\end{center}
\end{table*}

For the fast shocks $u_1>>u_*$ the amplified upstream magnetic
field may be estimated using the formula

\begin{equation}
\frac {B^2}{4\pi }\sim 2\eta _{esc}\rho u_1^2\sqrt{\frac {V_a}{c}}.
\end{equation}

For velocities $u_1>u_*$ Eq. (19) may be rewritten as

\begin{equation}
\frac {p_mc}{Z}=21\ \mathrm{TeV}\ \frac {\frac {\eta _{esc}}{0.05}
\left( \frac {u_1}{1000\
\mbox{km s}^{-1}}\right) ^2 n_H^{1/2}L_{pc} } { \ln \left( \frac
{2B_0}{B_b}\right) -1 +\left( 2\frac {u_1'^4}{u_3^4}-1\right)
^{1/4} } ,
\end{equation}
where $Z$ is the charge number of accelerated particles,
$u_3=1730$ km s$^{-1}$, $n_H$ is the hydrogen number density in cm$^{-3}$ and
$L_{pc}$ is the size $L$ in parsecs. Note that the denominator of this equation
contains the normalized shock velocity (12).

For three-dimensional shocks this equation may be considered
as a rough estimate of the maximum energy of accelerated
particles. Then the size $L$ is related to the shock radius $R$ and
the remnant age $T$  as $L=m_0R=u_1T$.

\section{Discussion}

The initiation of the MHD streaming instability by
accelerated particles in the shock precursor is an integral part of
the efficient diffusive shock acceleration process. The instability is
non-resonant if the electric current of accelerated particles is large
enough, so that the normalized shock velocity exceeds some critical
value

\[
u_1>(4cV_a^2/\eta _{esc})^{1/3}
\]
\begin{equation}=1340\ \mbox{km
s}^{-1}\left( \frac {V_a}{\mbox{10 km s}^{-1}}\right) ^{2/3}
\left( \frac {\eta _{esc}}{0.05}\right) ^{-1/3}.
\end{equation}
This equation is equivalent to Eq. (18) in Paper I.

The non-resonant character of this instability means that the
principal scale of the growing random magnetic field remains smaller
than the gyroradius of particles with momentum close to its  maximum
value $p_m(t)$.  The treatment of the non-magnetized particle scattering
is relatively simple and it allowed us to fulfill the
numerical simulation of the diffusive shock acceleration accompanied by
the strong streaming instability and to make the corresponding analytical
estimates. The calculated amplification of magnetic field in young SNR
proved to be very significant. The maximum energy of accelerated
particles is mainly limited by the finite time of the growth of
magnetic disturbances. The analytical estimate of the maximum
particle momentum is given by Eq. (19) and it is in agreement with our
numerical results presented in Fig.2. The resulting maximum momentum is
not as high as one may expect using the Bohm diffusion coefficient in
the amplified field since the scattering by the small scale field is
not so efficient.

The small-scale field approximation is broken
for particles with relatively low energies when
their gyroradii in
the amplified magnetic field are smaller than the principal scale
of the field.
Roughly it occurs when  the scattering
frequency of the particle is smaller than its gyrofrequency in the
amplified field. The ratio of these quantities for particles with
the maximum momentum $p_m$ is the ratio of the numbers from the 7th
and 3rd lines of Table 1. This ratio is close to 1 for the smallest
considered normalized velocity $u_1'=1550$ km s$^{-1}$ when the
small-scale approximation is only marginally valid. The ratio decreases if the
shock velocity increases. This ratio is close to 1/3
for the normalized velocity $4900$
km s$^{-1}$ that is the representative value for the young
historical supernovae (see below) . This
means that for this shock velocity the small-scale approximation
is valid for particles with the normalized momentum larger than
$p_m'/3$. For particles with smaller energies the amplified
magnetic field can be considered as the mean large-scale field.
These particles are resonantly scattered by the magnetic
inhomogeneities from the inertial range of the magnetic spectrum
(see Fig.4) or by the magnetic perturbations produced by the
resonant streaming instability of these particles (cf. Pelletier et al.
\cite{pelletier06}).

It is interesting that formula (21) without the root in the
denominator may be used to estimate the maximum energy of
particles, accelerated at slow astrophysical shocks when the
condition (22) is violated and the resonant streaming instability
should be taken into account. The matter is that the expression
for the increment of the non-resonant instability (17)
differs
from the increment of the resonant instability only by a factor
of the order of unity (see e.g. Berezinskii et al. \cite{berezinsky90}).
This is why the expression (21) may be also
used for slow shocks provided that the wave damping does not quench the
development of instability that is relatively slow in this case.
It may be in a highly ionized medium where the
damping of MHD waves on neutrals is negligible. Nonlinear damping
of Alfv\'en waves may be also depressed in a plasma with $\beta
>0.3$, since the presence of waves, propagating in the opposite to the cosmic
ray streaming direction is necessary for this damping
(Zirakashvili \cite{zirakashvili00}). Such waves may appear only
in a collisionless low-$\beta $ plasma  due to the nonlinear induced
scattering (Livshits \& Tsytovich \cite{livshits70}), or due to
the three-wave interactions in the framework of
magnetohydrodynamics (Chin \&
Wentzel \cite{chin72}).

Because the particles at the end of the spectrum are scattered by
the small-scale magnetic field,
the spectrum has
the universal shape in the cut-off region described by Eqs (7) and (8)
in the case of high velocity shocks.
This is important for
the calculation of gamma-ray production by the nucleon component in SNRs.

The comparison of the predicted values of amplified fields given in Table 1 with
the values derived from the thickness of X-ray filaments of young
SNRs (see e.g. V\"olk et al. \cite{voelk05}) shows reasonable
agreement, if the acceleration efficiency is not low: $\eta
_{esc}>0.05$.

The calculated downstream magnetic fields and
the maximum energies obtained for the  historical SNRs with known
ages are given in Table 2. We use the same supernova parameters as
 Vink \cite{vink06}.  We
assumed the standard interstellar value of the magnetic field
$B_0=5\ \mu $G for SNRs Kepler, Tycho and SN1006. The accepted
magnetic field strength $B_0=10\ \mu $G for Cas A supernova
gives the value of the Alfv\'en velocity about 10 km s$^{-1}$ that
is a  reasonable number for the stellar wind produced at the Red Supergiant stage
of the likely progenitor of this supernova. The
magnetic field amplified upstream of the shock is determined by
the value of the normalized shock velocity $u_1'$ and is given in
the  3rd line of Table 1. This almost isotropic random magnetic
field is further amplified by a magnetic compression factor $\sigma _B=4$ in the
shock transition region.

One of the key parameters in our calculations $\eta _{esc}$ is in
principle determined by the injection efficiency of thermal particles
in the process of acceleration and by the degree of shock
modification by the cosmic ray pressure, see Berezhko and Ellison
\cite{berezhko99} and Blasi et al. \cite{blasi05}.
The value of $\eta _{esc}$ can not be
calculated theoretically yet and we used different $\eta _{esc}$ in our
estimates.
The results
obtained for three values $\eta _{esc}=0.01$, $\eta _{esc}=0.05$ and $\eta _{esc}=0.14$
are shown
in Table 2. The value $\eta _{esc}=0.14$ is realized for
the plain cosmic ray modified shock with the compression ratio $\sigma =6$
and the thermal sub-shock compression ratio $\sigma _s=2.5$. The second
value $\eta _{esc}=0.05$ corresponds to the situation when due to some
reason the shock modification is weaker. The value $\eta _{esc}=0.01$ corresponds
to the non-modified shock with the cosmic ray pressure $P_{\mathrm{CR}}$ about
ten percents of the ram pressure $\rho _0u_1^2$.

The maximum energies of particles accelerated in Kepler, Tycho and SN1006 supernovae that all
are of the Ia type are
about $100\div 300$ $Z$ TeV. These maximum energies
were not strongly different in the past during the free
expansion stage of the remnant evolution
since the higher
shock velocity was almost compensated by the smaller shock radius in Eq. (21).

The similar maximum energies are predicted for the core collapse
IIP supernovae. They have large ejected masses about $M_{ej}\sim 10
M_{\odot }$ and correspondingly relatively low expansion
velocities of the order of $3000$ km s$^{-1}$ (see Chevalier \cite{chevalier05} for a review).

The situation is different
and the maximum particle energies can be higher
in the case of the core collapse Ib/c and IIb supernovae. These
supernovae have
high initial velocities about $3\cdot 10^4$ km s$^{-1}$, small
ejected masses $M_{ej}=1\div 3\ M_{\odot}$
and the circumstellar medium corresponding to the dense red supergiant
stellar wind (IIb supernovae ) or the interaction zone between
the fast stellar wind of a Wolf-Rayet progenitor and the slow red supergiant wind (Ib/c supernovae)
with strong magnetic fields (see Chevalier \cite{chevalier05} for a review).
In the present paper we
have considered the acceleration of
particles and the generation of the MHD turbulence
 at the  parallel shock and  our theory can not be directly applied
 to this type of supernovae because the
magnetic field in the stellar wind is azimuthal. However, it is possible
that stellar winds contain
significant  random magnetic fluctuations and we may expect
that some part of a SNR shock surface may be
treated as a parallel shock.
This also will provide the injection of thermal particles into the diffusive shock
acceleration process since it is known that the injection of
thermal ions occurs preferentially
at the parallel shocks (see V\"olk et al. \cite{voelk03} for
discussion of these topics).

The Cas A is a  Ib type supernova. The maximum energy was larger
in the past because the small radius of the shock in Eq. (21) was
compensated by the higher gas density of the  stellar wind and
the higher shock velocity. Thus the velocity
$u_1=10^4$ km s$^{-1}$ gives the maximum energy $p_mc=1.2$ $Z$ PeV.
Therefore such supernovae can accelerate cosmic ray protons up to PeV energies.

Because the maximum momentum (21) decreases at the Sedov stage when the
age of the remnant increases, the highest energy particles leave
the remnant. The overall spectrum produced by the SNR is formed in
this manner (Ptuskin \&
Zirakashvili \cite{ptuskin05}) with the energy spectrum close to
$E^{-2}$. At the earlier free expansion stage when only a small
amount of the supernova ejecta energy is transferred to the
supernova shock the steep high energy tail in the spectrum is
formed ( Berezhko \& V\"olk \cite{berezhko04}, Ptuskin \&
Zirakashvili \cite{ptuskin05}). This means that the particles from the end of the cosmic
ray spectrum produced by the given supernovae are accelerated at
the end of the free expansion stage when the shock velocity is of
the order of the characteristic ejecta velocity. It is about
$7000\div 10000$ km s$^{-1}$ for Ia/b/c and IIb supernovae.

We should note that the shock velocities in Table 2 are based on radio and
X-ray expansion measurements. If we use the lower shock velocities
from V\"olk et al. \cite{voelk05}, the corresponding maximum energies
are a factor of 2 smaller.

It is clear from Fig.7 that the amplified magnetic field does not
drop downstream of the shock. Its level is maintained by the turbulent
motions produced by the interaction of density inhomogeneities with
the shock front. However, the amplitude of these motions slowly
decreases downstream of the shock. At some distance magnetic
and kinetic energies will become equal to each other. At larger distances
magnetic dissipation may occur. These distances are larger than $2\div 3L_z$
according to Fig.7. We conclude that the dissipation length of the magnetic
field downstream of the shock is not smaller than $0.1L$ in our numerical
simulation.

The interaction of the shock front with density disturbances results in the
shock front deformation (see Fig.6). The thickness of X-ray rims produced
by the synchrotron cooling of accelerated electrons should
increase correspondingly up to the values about 0.01$L$ according to
Fig.7 (a so-called projection effect is not taken into account here).
Probably this is the reason for the relatively
small value of the magnetic field $485$ $\mu $G in Cas A
inferred from the width of X-ray filaments
in comparison with the theoretical value expected at $\eta _{esc}>0.05$,
see Table 2.

The characteristic time of the random motions of the shock front
is given by the ratio of the size of density inhomogeneities and the shock velocity.
It is about one year for the size of density inhomogeneities $10^{16}$ cm
in young historical SNRs. The value $L=0.5R=1.5$ pc was assumed for this estimate.

PDF of the magnetic field downstream of the shock
that is shown
in Fig.9  is described with a good accuracy by the
following function:

\begin{equation}
P_B(B)=\frac {\sqrt{6}B}{\left< B^2\right> }\exp
{\left( -\sqrt{6}B/\left< B^2\right> ^{1/2}\right) }.
\end{equation}
The exponential tails of the magnetic PDF appear to be a fairly
universal feature of turbulently amplified magnetic fields
(Brandenburg et al. \cite{brandenburg96}, Schekochihin et al.
\cite{schekochihin04}).

As one can see from Table 1, the electric potential $\phi $ has rather
large values for the fast shocks with velocities $u_1'>30000$
 km s$^{-1}$. The mean electric field is directed
opposite to the direction of the diffusive electric current. This
 electric field drags the particles which produce the instability
toward the shock. It may create the steepening of the spectrum
of these particles. If there is a small amount of oppositely
charged particles (electrons), their spectrum should be somewhat
flatter.

\section{Conclusion}

We have investigated the acceleration of particles at the fast
plane parallel shock. The generation of MHD turbulence by the
non-resonant streaming instability (Bell \cite{bell04}) was taken
into account. We solved this problem using only first
principles and with a minimum of simplifying assumptions. We
combined the analytical solution for the particle acceleration and
the numerical MHD calculations for the evolution of the MHD turbulence.
The following results were obtained:

1) For the relatively fast shocks when the  condition (22) is satisfied,
the particles at the high-energy end
of the spectrum are scattered by small-scale random magnetic fields
generated by the non-resonant streaming instability.
Their spectrum has the universal shape given by Eq. (8).

2) The MHD turbulence is mainly generated by the streaming of run-away
particles at large distance from the shock. The level of the MHD
turbulence is the  highest in the shock vicinity. The accelerated
particles are concentrated in the same region. This means that the
acceleration of particles can be considered in the one-dimensional approximation
even for a three-dimensional system. The characteristic width of
the particle distribution in our simulations is not larger than $0.1\ L $
(see Fig.3 and 10th
line of Table 1).

3) It is important, that the non-resonant instability produces
 strong density fluctuations upstream of the shock (see Sect.4 and Paper I).
These fluctuations produce a strong deformation of the shock
front and fast vortex motions downstream of the shock. That is
why the magnetic amplification in the shock transition region is
not reduced to a simple compression of the magnetic field in the
direction perpendicular to the shock front. The magnetic field is
also stretched by the flow motions in the direction perpendicular to the shock
front. As a result, the magnetic field component which is
perpendicular to the shock front is a factor of 1.4 larger than
the parallel components downstream of the shock. This naturally
explains the preferable radial orientation of magnetic fields in
young SNRs.

The characteristic time 1 year of the shock deformation and of the corresponding MHD
fluctuations found here (see Sect.6) is of particular interest in
the light of the last results on variability of X-ray emission
observed in RXJ1713 SNR (Uchiyama et al. \cite{uchiyama07}).

4) The dissipation of the magnetic field downstream of the shock is
 relatively slow in our simulation. If so, the origin of X-ray filaments, observed
 in young SNRs, is related to the fast synchrotron cooling of accelerated
 electrons but not to the decay of the MHD turbulence.

5) The values of the calculated amplified magnetic field are similar to those,
observed in historical SNRs, if the energy flux of the run-away
particles is not low: $\eta _{esc}>0.05$ (see Table 2).

6) The magnetic field growth is only linear in time for the
fast shocks with the normalized
velocities higher
than about ten thousand km s$^{-1}$.
This reduces the maximum energy of accelerated particles
compared to the case of the exponential growth.
 For these shocks the energy
of the magnetic field amplified upstream is a small fraction $\sqrt{V_a/c}$ of
the energy density of the highest energy particles (see Eq.(20)).
The magnetic field
amplification is relatively weak for slow shocks with the normalized
velocities 1230 km s$^{-1}$ and smaller (see Eq.(18)).

7) We calculated numerically the maximum energy of accelerated particles
 (see Fig.2). This energy may be described by the analytical
formula (19). The maximum energies of particles are higher than
the energies obtained in the Bohm limit in the background magnetic field
(see the 7th line of the Table 1) but lower than the energies
obtained using the Bohm limit in the amplified field.
The significantly long
time of the random magnetic field growth
is the main  factor that limits the maximum energy of accelerated particles.
We calculated the
maximum energies for four historical SNRs (Table 2).

8) Using the last result we found that the maximum energy of
cosmic ray protons accelerated by Ia and IIP supernovae is about
$100\div 300$ TeV. Only Ib/c and presumably IIb supernovae may
accelerate protons up to PeV energies. Since it is expected that
the explosion rate is the highest for IIP supernovae, we should
observe a change of the slope of the galactic cosmic ray
spectrum at the energies of the order of 100 TeV.

9) The MHD turbulence generated by the non-resonant streaming
instability has a non-zero magnetic helicity. The helical Lorentz
force produces corresponding plasma motions and the mean electric
field ${\bf E}_0$ that is in the opposite direction to the electric
current of energetic particles. This electric field modifies the
cosmic ray transport equation (see Paper I). This effect is
significant for very fast shocks with velocities larger than 30
thousands km s$^{-1}$. The presence of this field should result in the steepening of the
spectrum of particles which produce the  non-resonant instability
(presumably nucleons) and the flattening of the spectrum of
oppositely charged particles (presumably electrons).

\begin{acknowledgements}
We thank the anonymous referee for a number of valuable suggestions.
We are grateful to Heinz V\"olk for many fruitful discussions of acceleration
by astrophysical shocks.
VNZ and VSP  acknowledge the hospitality of the
Max-Planck-Institut f\"ur Kernphysik, where this work was mainly carried
out. The work was also supported by the RFBR grant in Troitsk.

\end{acknowledgements}
\appendix

\section{The dependence of scattering on pitch angle. }

If the random field is not isotropic the same is true for the particle
scattering. Let us
assume that the random magnetic field is isotropic in the plane perpendicular to
the mean magnetic field ${\bf B}_0$ and that the cosmic ray distribution $f_0$
depends only on pitch
angle $\theta $. Now the scattering operator described by the tensor $\nu _{ij}$ (cf. Paper I)
can be written as

\begin{equation}
\frac \partial {\partial p_i}\nu _{ij}\frac {\partial f_0}{\partial p_j}=
\frac {\partial }{\partial \mu }\frac {\nu _0(\mu )}{2}(1-\mu ^2)
\frac {\partial f_0}{\partial \mu } ,
\end{equation}
where $\mu =\cos{\theta }$. The scattering frequency $\nu _0(\mu )$ can be expressed in terms of
the spectrum of $y$-component of the random magnetic field
(mean field is in $z$ direction):

\begin{equation}
\nu _0(\mu )=
2\pi \frac {q^2v}{p^2c^2}\int d^3k B_{yy}({\bf k})\delta (k_z\mu +k_x\sin{\theta }) .
\end{equation}
The parallel diffusion coefficient $D_{\parallel }$ can be
written now as (see e.g. Berezinskii et al. \cite{berezinsky90})
\begin{equation}
D_{\parallel }=\frac {v^2}{2}\int \limits ^1_0d\mu \frac {1-\mu
^2}{\nu _0(\mu )}.
\end{equation}

\end{document}